\documentclass[twocolumn,showpacs,fleqn,nobibnotes]{revtex4}
\usepackage{amsmath}
\usepackage{graphicx}
\usepackage{float}
\usepackage{subfigure}
\usepackage{verbatim}
\usepackage{color}
\usepackage{comment}
\usepackage{hyperref}

\usepackage{booktabs}

\begin{document}
	
	\title{Central exclusive $\chi_{c,b}$ production at high energy colliders and gluon saturation approach}
	\author{F. Kopp, M.B. Gay Ducati, M.V.T. Machado}
	
	\affiliation{High Energy Physics Phenomenology Group, GFPAE  IF-UFRGS \\
		Caixa Postal 15051, CEP 91501-970, Porto Alegre, RS, Brazil}

\begin{abstract}
The central exclusive production of $\chi_{c}$ and $\chi_{b}$ at the LHC, RHIC and Tevatron energies is analyzed, using the recent unintegrated parton distribution (UGDs) functions available in the package TMDlib. Comparison with data is performed, which tests the underlying assumptions basing the theoretical approach and  it can constrain the unintegrated gluon distribution function at the small-$x$ region. Predictions for LHC energies using recent UGDs based in CCFM formalism are provided. It is explored the  underlying uncertainties on this production as the choice for the unintegrated gluon distribution and factorization scale is done. Moreover, based on the parton saturation model for the gluon distribution, analytical expressions for the rapidity distributions are proposed. The prompt production of $J/\psi+\gamma$ and $\Upsilon+\gamma$ is computed for the first time for LHC energies within the very same formalism used for $\chi$ production.   
\end{abstract}	

\maketitle
	
\section{Introduction}	
The central exclusive production (CEP) processes are considered as an useful way for testing perturbative and nonperturbative aspects of QCD \cite{Albrow:2010yb}. CEP is a process where the incident hadrons remain intact after the interaction, and an additional simple central system is
produced. In Regge language,  CEP allows us to study the structure of Pomeron since we have a double Pomeron exchange leading to a specific final state like Higgs boson \cite{GayDucati:2008zs,GayDucati:2010xi}, scalar and tensor mesons including charmonium states as $\chi_{c,b}$ mesons.  It carries the quantum numbers of the vacuum, so it is a colorless entity in QCD and reactions initiated by Pomerons are experimentally characterized by the rapidity gap events. At the Large Hadron Collider (LHC), investigations on CEP are very active \cite{Ewerz:2019arb,N.Cartiglia:2015gve}. Our focus in this study is the production of $\chi_c$ and $\chi_b$ within the two-gluon exchange formalism and the non-relativistic approach for evaluating the $P$-wave quarkonium decays \cite{Yuan:2001nu}. At the LHC, the LHCb Collaboration has done analyses of CEP of $\chi_c$ mesons, reconstructed in the decay $\chi_c \rightarrow J/\psi \gamma \rightarrow \mu^+\mu^-$ \cite{SantanaRangel:2019gob,Volyanskyy:2013qco,LHCb:2011dra} at 7 and 8 TeV. The measured cross sections times branching ratios $\chi_c$ states reach to dozens of picobarns  \cite{Ewerz:2019arb,N.Cartiglia:2015gve}, which is in agreement with theoretical predictions \cite{LHCb:2011dra}. Search for the CEP of $\chi_b$ mesons has been done, however the background contributions are not completely determined. In any case, most theoretical predictions for the cross section for $\chi_b$ give values lower than 1 fb which implies very few events. The ALICE Collaboration has recorded zero bias and minimum bias data in proton-proton collisions at a center-of-mass energy of $\sqrt{s}=7$ TeV. Events containing double gap topology have been studied and they are associated to CEP \cite{Schicker:2014wvk,Schicker:2019qcn}. In particular, central meson production was observed and it was verified that  $K_s^0$ and $\rho^0$ are highly suppressed while the $f_0(980)$ and $f_2(1270)$ with quantum numbers $J^{PC}=(0,2)^{++}$ are much enhanced (one of us calculated the CEP of these $f_0$ and $f_2$ mesons in Ref. \cite{Machado:2011vh}) . The measurement of those states is an evidence that the double gap condition used by ALICE selects events dominated by CEP or related  processes. ATLAS and CMS also measured inclusive $\chi_{c,b}$ production and cross section ratios for different states are studied \cite{ATLAS:2014ala,Chatrchyan:2012ub}. Further program on CEP is ongoing in both collaborations \cite{Tasevsky:2017wne,Khakzad:2017mmd} and also at the Relativistic Heavy Ion Collider (RHIC) \cite{Adamczyk:2014ofa}.

On the theoretical side, applications to $\chi_c$ production is considered by the Durham group in \cite{Harland-Lang:2014lxa} and references therein, including perturbative QCD and also a non-perturbative component. In Ref. \cite{Yuan:2001nu} it was calculated the $\chi_{c}$   and $\chi_b$  CEP cross sections for the Tevatron, in the forward approximation neglecting the $\chi_{c1}$ and $\chi_{c2}$ states.  The Bialas-Landshoff model is applied to $\chi$ meson production in Ref. \cite{Bzdak:2005rp}, consistent with the $\chi_{c0}$ cross section found by SuperCHIC MonteCarlo for the same parameters \cite{Harland-Lang:2015cta,Harland-Lang:2018iur}. The Cracow/Lund group performed calculations \cite{Pasechnik:2007hm,Pasechnik:2009bq}, using a different approach than the Durham group for the unintegrated gluon distributions (UGDs), ${\cal{F}}_g$ , and taking into account Quasi Multi-Regge Kinematics for the subprocess vertex. The cross sections are found to have a large dependence on
the model parameters and the choice of gluon distributions. Interestingly, the cross sections vary by an order of magnitude when using distinct UGDs.
 
 The focus of this work is the central exclusive production of heavy quarkonium $(\chi_c,\chi_b)$ at the Tevatron and at the
LHC. One motivation for this study is that $\chi$-production probes the gluon density down to fractional gluon momenta of $x\sim 10^{−6}$, being potentially sensitive to saturation
effects. Moreover, this kind of exclusive process is a standard candle in QCD calculations and brings information on the off-forward unintegrated gluon distribution.  These objects are poorly
constrained in the kinematics investigated here and it is a timely investigation. The formalism of Ref. \cite{Yuan:2001nu} is considered taking into account the new fitted UGDs available in TMDlib (Transverse Momentum Dependent parton distributions) package\cite{tmdlib}. The UGD's used here were those based on CCFM model with three different fitting sets. In addition we consider an analytical UGD based on parton saturation model, i.e. the celebrated GBW saturation model \cite{GolecBiernat:1999qd}. The purpose to use UDG's based on GBW model was to quantify the deviation of a simple model from a robust model like CCFM and investigate the role played by saturation physics in the UGDs at high energies. The novelty of the results is the updated computation of cross sections using the last CCFM-based UGDs and the predictions for the prompt production of $J/\psi+\gamma$ and $\Upsilon +\gamma$ in the very same formalism. We provide analytical expressions for rapidy distributions for prompt $\chi$ and $V+\gamma$  production, Eqs. (\ref{gbwfull}) and (\ref{xs-VgGBW}), based on QCD parton saturation which are quite useful for further phenomenological studies. For the first time the estimation of nuclear effects are predicted for $pA$ and $AA$ collisions within the geometric scaling  approach, shown in Eqs. (\ref{Adep}). This is crucial for LHC analyses, where the nuclear saturation scale, $Q_{s,A}^2(y) \approx A^{4/9}(10^{-5}\sqrt{s}/m)^{\lambda}e^{\lambda y}$ (with $\lambda \simeq 0.25$) , should be close or larger than the meson mass for a given forward rapidity $y$.

 This paper is organized as follows. In the next section the theoretical formalism is presented, including the main building blocks and the relevant parameters. In Sec. \ref{sec3}, results of the calculations are presented and we compare them with the current literature. In the last section we summarize our main conclusions and remarks.

\section{Theoretical Formalism}
The central exclusive $\chi$ production, $p+p(\bar p)\rightarrow p+\chi_J+p(\bar p)$, is analyzed in the two gluon exchange model \cite{Yuan:2001nu}, where the hard 
sub-process $gg\rightarrow \chi_J$ is initiated by gluon-gluon fusion and the 
second $t$-channel gluon (with transverse momentum $k_{\perp}$) 
is needed to screen the color flow across the rapidity gap intervals. For the hadronization, a non-relativistic approach is used to compute the $P$-wave quarkonium decays. Given the forward scattering amplitude, ${\cal M}$, the rapidity distribution of $\chi$ production will be
\begin{equation}
\label{xs-ch0}
 \frac{d\sigma}{dy}=\int \frac{|{\cal M}|^2}{16^2\pi^3}e^{Bt_1}e^{Bt_2}dt_1dt_2,
\end{equation}
where $y$ is the rapidity of the $\chi$ state. Moreover, $t_i$ is the momentum transfer squared at the proton (anti-proton) vertices, 
and $B$ is slope for the proton form factor, which will be taken as 
$B=4.0\,\mathrm{GeV}^{-2}$. After integrating on momentum transfer $t_1$ and $t_2$, one obtains \cite{Yuan:2001nu},
\begin{eqnarray}
\label{xs-ch}
\frac{d\sigma}{dy} &= & S^{2}\frac{\pi^4\alpha_s^2 m_{\chi}}{B^2}|R'_P(0)|^2 \,I_g^2, \\
I_g & = & \int \frac{dk_{\perp}^2}{(k_{\perp}^2)^2}
\frac{{\cal{F}}_g(x_1,x_1',k_{\perp},\mu^2){\cal{F}}_g(x_2,x_2',k_{\perp},\mu^2)}{(m_{\chi}^2+k_{\perp}^2)^2}, \nonumber
\end{eqnarray}
where ${\cal{F}}_g$ are the unintegrated off-forward (skewed) gluon distribution
functions, computed at a perturbative scale $\mu^2$. For the masses and first derivative of radial $P$-wave functions \cite{Eichten:2019hbb}, we use $m(\chi_{c0})=3.414$  GeV with $|R'_P(0)|^2_c=0.075$ GeV$^5$ and $m(\chi_{b0})=9.859$ GeV with $|R'_P(0)|^2_b=1.42$ GeV$^5$ (notice that the recent values for wave functions 0.1296 and 1.6057 will increase cross section by a factor 1.73 and 1.13, respectively) . The rapidity gap survival factor $S^2$ for central exclusive   $\chi_J$ production can be calculated using the formalism of \cite{Harland-Lang:2014lxa}, which gives: 
\begin{equation*}
S^{2}(\text{Tevatron})=0.058 \ , \
S^{2}(\text{LHC})=0.029.
\end{equation*}

Regarding the UGDs, they can be obtained from the conventional gluon density as \cite{Harland-Lang:2014lxa}
\begin{eqnarray}
{\cal{F}}_g(x,x',k_{\perp},\mu^2)=R_g\frac{\partial}{\partial {\rm ln} k_{\perp}^2}
\left[\sqrt{T(k_{\perp},\mu^2)}xg(x,k_{\perp}^2)\right],
\end{eqnarray}
where the factor $R_g$ takes into account the skewed effects of the off-forward gluon density compared with the conventional gluon density in the region of $x'\ll x$. The factor $T^2$ \cite{Harland-Lang:2014lxa} will reduce to 
the conventional Sudakov form factors in the double logarithmic limit. The R$_{g}$ factor used in the literature are $ R_{g}(\text{Tevatron}) = 1.4 \ , \ 
 R_{g}(\text{LHC}) = 1.2$. This factor produces a factor equal to 3.84 (Tevatron) and 2.07 (LHC), since it appears as  R$_{g}^{4}$ in Eq. (\ref{xs-ch}). However, we used R$_{g}$=1 in order to compare predictions to others works in literature.

We will consider here two implementations of UGDs. The first one  is the new fitted UGDs available in TMDlib (Transverse Momentum Dependent parton distributions) library \cite{tmdlib}, based on CCFM model with three different fitting sets. In this case, in the numerical calculations we used the $\alpha_s(m_{\chi}^{2})$ to a one-loop order (LO) and 4-flavors ($n_f=4$). For each CCFM set it was used a specific $\Lambda_{QCD}$, using the prescription given in Ref. \cite{tmdlib}. The second considered UGD is taken  from the saturation model \cite{GolecBiernat:1999qd}, which is analytical and with parameters fitted from DIS data at small-$x$. It reads as,
\begin{eqnarray}
{\cal{F}}_g(x,x', k_{\perp}) =R_g \frac{3\sigma_0}{4\pi^2\alpha_s}\left(\frac{k_{\perp}^4}{Q_s^2}\right)\exp\left(-\frac{k_{\perp}^2}{Q_s^2}\right), 
\label{GBWUGD}
\end{eqnarray}
where $Q_s(x)=(x_0/x)^{\lambda/2}$ is the saturation scale giving the transverse momenta transition between the dilute and saturated gluon system. It presents the geometric scaling property, i.e. the UGD depends on the scaling variable $k_{\perp}^2/Q_s^2(x)$ and not separately on $x$ and $k_{\perp}$. In the numerical calculation, the updated values for the model parameters (fit result including charm) were used:  $\sigma_0=27.32$  mb, $\lambda = 0.248$  and $x_0=4.2\times 10^{-5}$   \cite{Golec-Biernat:2017lfv}. Also, at large rapidities we multiply the GBW UGD by the large $x$ threshold, $(1-x)^5$.

In the next section, the uncertainties on theoretical predictions are investigated and a closer look in the parton saturation model is applied to the CEP of quarkonium.

\section{Results and discussions}
\label{sec3}

Here, a focus on the exclusive production of mesons $\chi_{c,b}$ in proton-proton collisions at LHC energies is taken. The present investigation is relevant for the ATLAS, CMS and ALICE experiments. The theoretical formalism presented in previous sections and  its theoretical uncertainties will be investigated. In particular, the uncertainty coming from the choice for the unintegrated gluon distribution taking into account  different prescription for the renormalization/regularization scale $\mu^2$. As a cross check, predictions are performed also for the lower energy at the  Tevatron. The distribution for the meson rapidity is presented and for completeness it is computed the corresponding integrated cross sections.

\subsection{Unintegrated gluon distribution}
\begin{figure}[t]
		\includegraphics[scale=0.4]{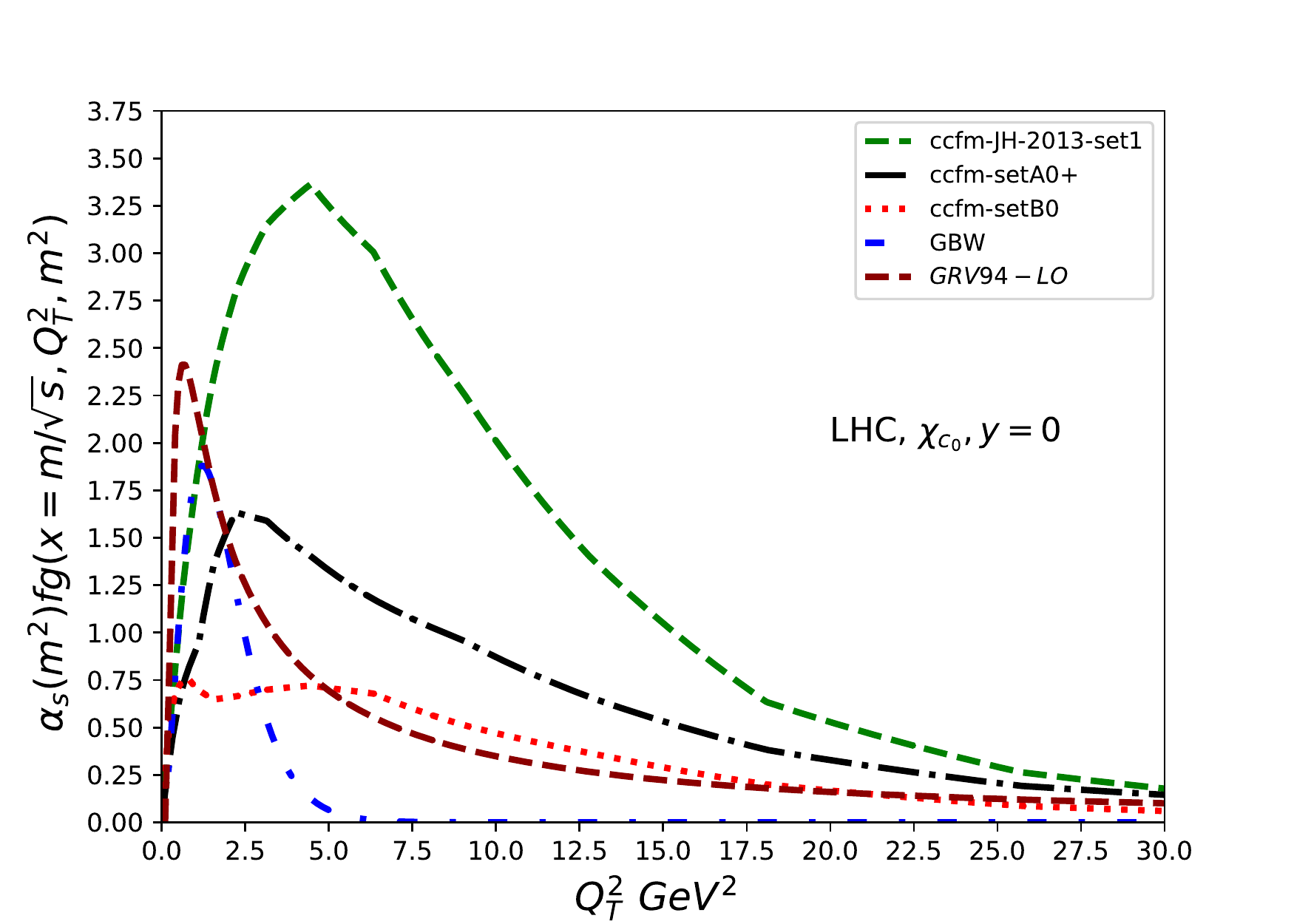} 
	\caption{Unintegrated gluon distributions as a function of $k_{\perp}^2$ at scale of $\mu^2=\chi_{c0}^2$ for the different sets of UGD's at the LHC energy. }
	\label{fig:udgmchi02}
\end{figure}

In this section the different sets of UGD's for distinct choices for the hard scale are compared. Namely, it is investigated the role played by $\mu^2$, using the prescriptions $m_{\chi}^2/4\leq \mu^2 \leq m_{\chi}^2$. Starting with the $\chi_{c0}$ production, on  Fig. \ref{fig:udgmchi02} is shown the behavior on transverse momentum, $k_{\perp}^2$, for different sets of the gluon distribution at central rapidity, $y=0$, at 14 TeV (LHC energy). At midrapidities the typical gluon momentum fraction is $x_1=x_2\sim 10^{-4}$ with a not so hard scale $3< \mu^2 < 11$ GeV$^2$. In this kinematic range, parton saturation physics (taming the gluon distribution at small-$x$) could be important \cite{Ayala:1996ed,Ayala:1996em,AyalaFilho:1997du}. Three sets for CCFM UGD are presented (JH-2013-set1, set A0+ and set B0), as well as the gluon saturation UGD from GBW model and the UGD from GRV94-LO. It can be seeing that the peak occurs for larger $k_{\perp}^2$ in CCFM compared to GRV94 and GBW UGD's. This is directly related to the starting scale $Q_0^2$ in hard scale evolution and the extrapolation for small gluon transverse momenta. For UGD's extracted from parton saturation physics, the peak occurs around the saturation scale, $Q_s^2\sim (x_0/x)^{0.3}$ (with $x_0\simeq 10^{-4}$). Therefore, at central rapidity at the LHC the saturation scale is of order $Q_s^2 \simeq 1$ GeV$^2$, which is confirmed by the numerical results. All results shown in Fig. \ref{fig:udgmchi02} are computed with $\mu^2=m_{\chi_c}^2$.

\begin{figure}[t]
		\includegraphics[scale=0.4]{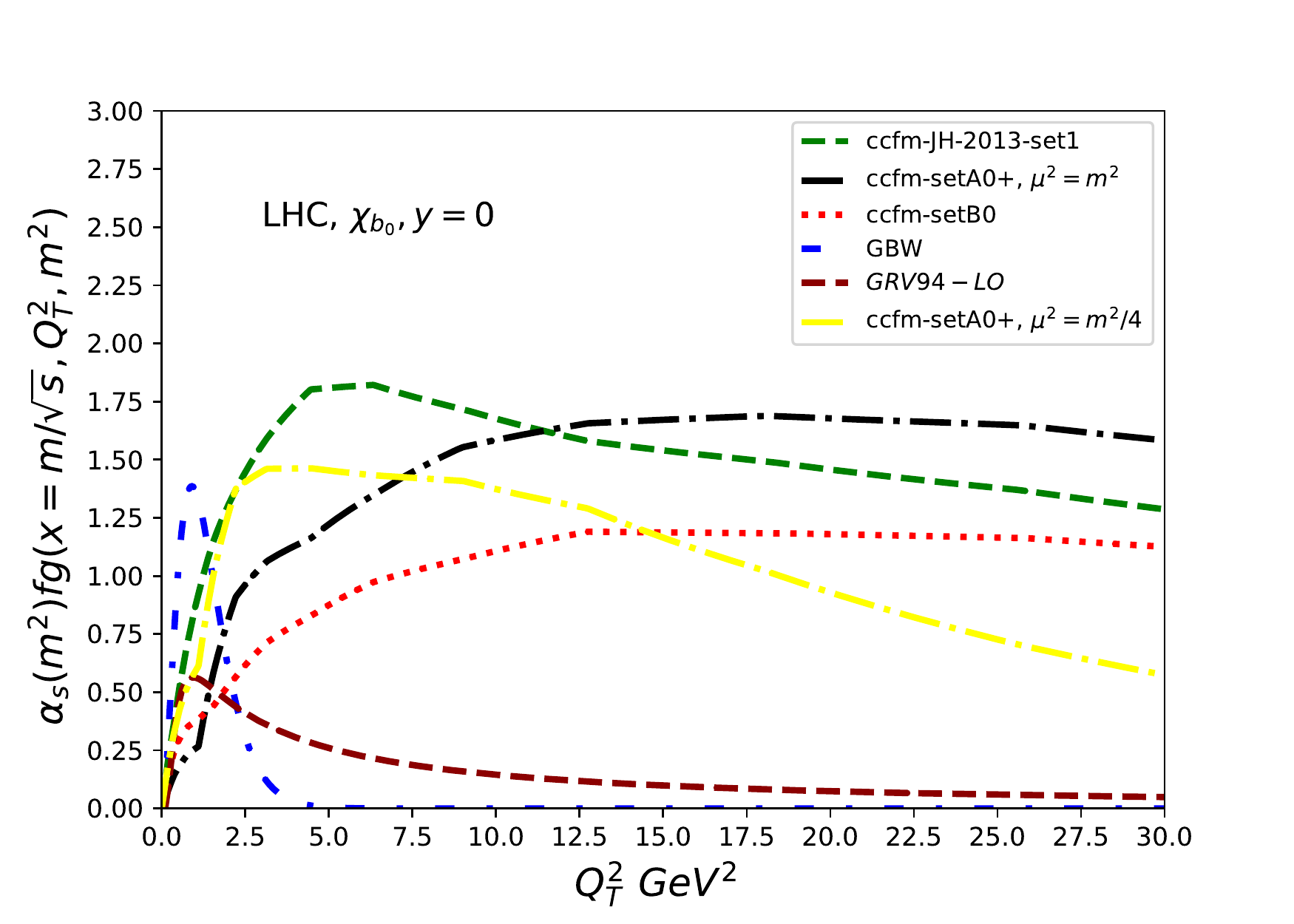} 
	\caption{Unintegrated gluon distribution as a function of the gluon transverse momentum at scale $\mu^2=\chi_{b0}^2$ for different sets of UGD's at the LHC energy. For set CCFM A0+, we also compute the UGD at scale $\mu^2=\chi_{b0}^2/4$. }
	\label{fig:udgchiblhc}
\end{figure}

We turn the corresponding analyses for $\chi_{b0}$ production. In this case, at midrapidity gluons have $x_{1,2}\sim 10^{-3}$ probed at scale $25< \mu^2 < 100$ GeV$^2$. In order to single out the uncertainty related to the choice of the hard scale, the sets A0+ taking $\mu^2=m_{\chi_b}^2$ and $\mu^2=m_{\chi_b}^2/4$ are compared. This is shown in Fig. \ref{fig:udgchiblhc}, using same notation as the previous figure. Basically, it is found that for a larger scale the contribution from the gluon with large transverse momenta is increasingly important. Once again, the GBW UGD is peaked near the saturation scale and large transverse momenta contributions are exponentially suppressed. In what follows,  the numerical results for the sets we have discussed above are investigated.

\subsection{Differential cross section}

\begin{figure}[t]
		\includegraphics[scale=0.45]{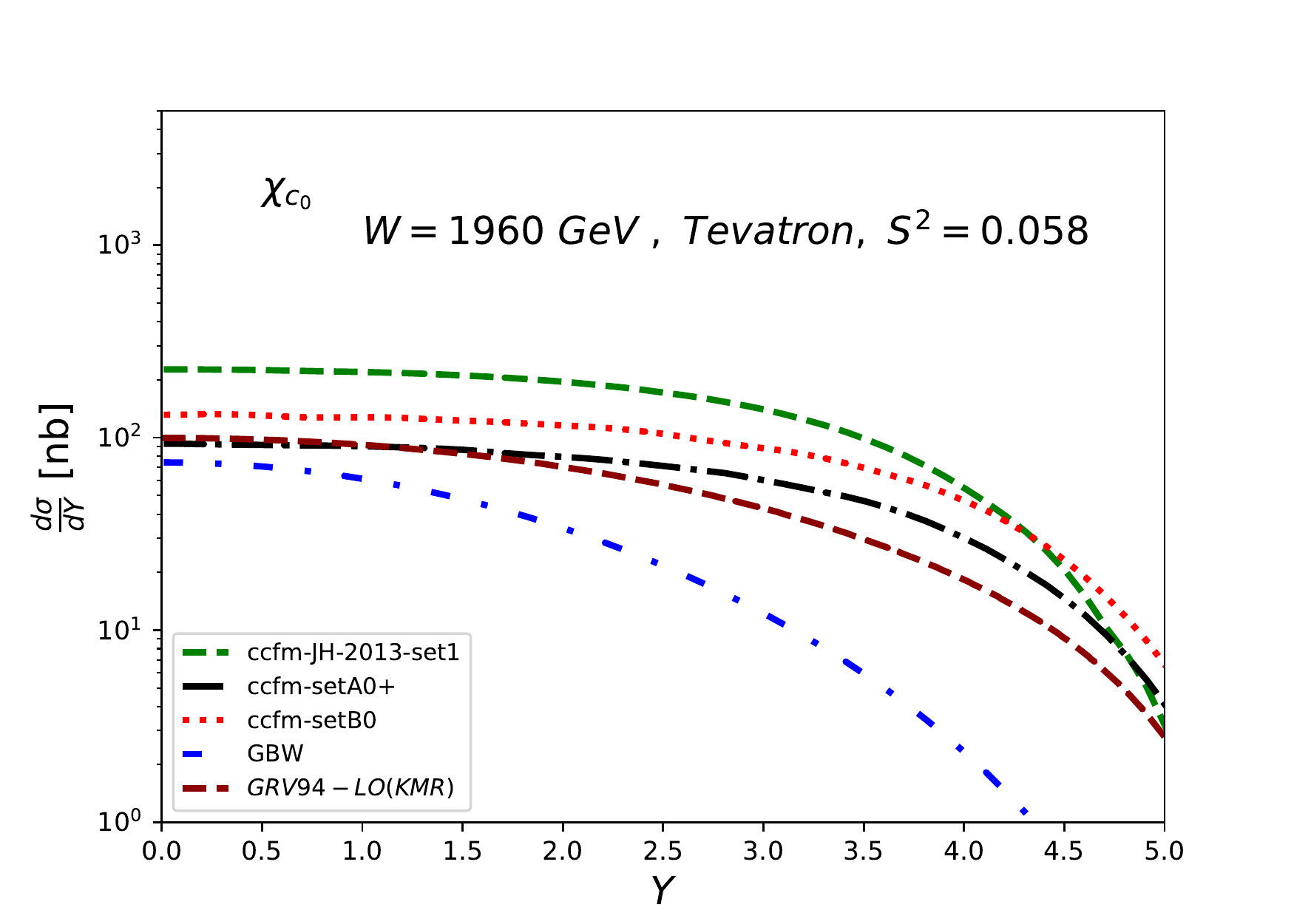} 
	\caption{The rapidity distribution for exclusive $\chi_{c0}$  production using four different sets of UGD's at Tevatron energy. }
	\label{fig:chic0tevatron}
\end{figure}

\begin{figure}[t]
		\includegraphics[scale=0.45]{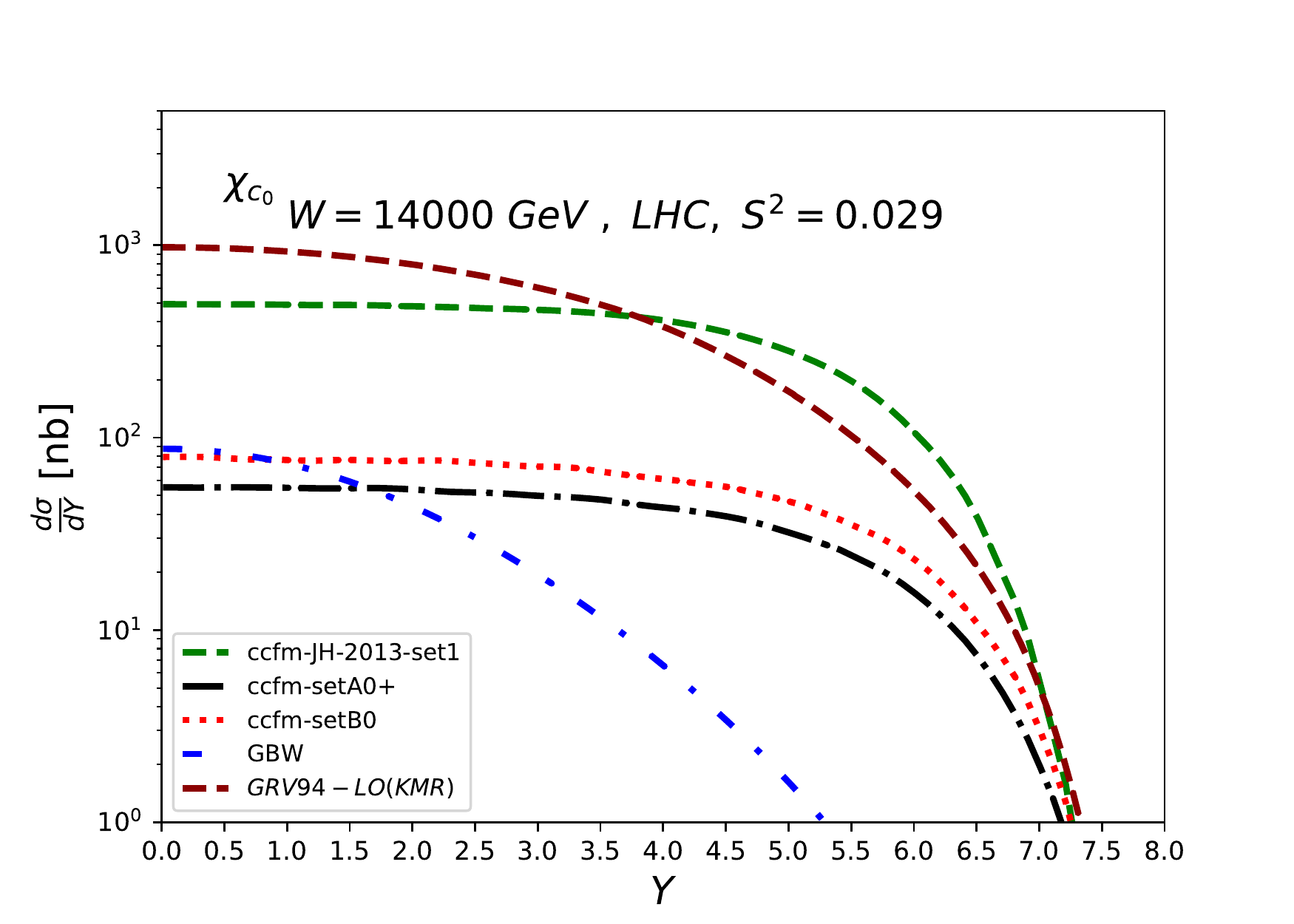} 
	\caption{Rapidity distribution for exclusive $\chi_{c0}$ production at the LHC. }
	\label{fig:chic0lhc}
\end{figure}

The predictions for the rapidity distribution, $y$, for the exclusive $\chi_{c,b}$  production are obtained, and for sake of completeness a cross check for Tevatron energies,  shown in Fig.
 \ref{fig:chic0tevatron} (the curve label is the same as the  previous figures)  was done. Here, $\mu^2=m_{\chi_{c0}}^2$. The behavior is similar for different sets except for the GBW UGD. The suppression at large rapidities compared to CCFM and GRV94-LO is evident and this trend is more dramatic for LHC energies. The predictions for LHC are presented in Fig. \ref{fig:chic0lhc}, where the choice for distinct sets for UGD's leads to one order of magnitude difference at midrapidities. This can be traced out to the uncertainty on the determination of  the gluon distribution  at very small-$x$. One has $\frac{d\sigma}{dy}(y=0)\sim 100$ nb for Tevatron and $\frac{d\sigma}{dy}(y=0)\sim 300$ nb, with $R_g=1$, and a sizable spread for LHC case.

\begin{figure}[t]
		\includegraphics[scale=0.45]{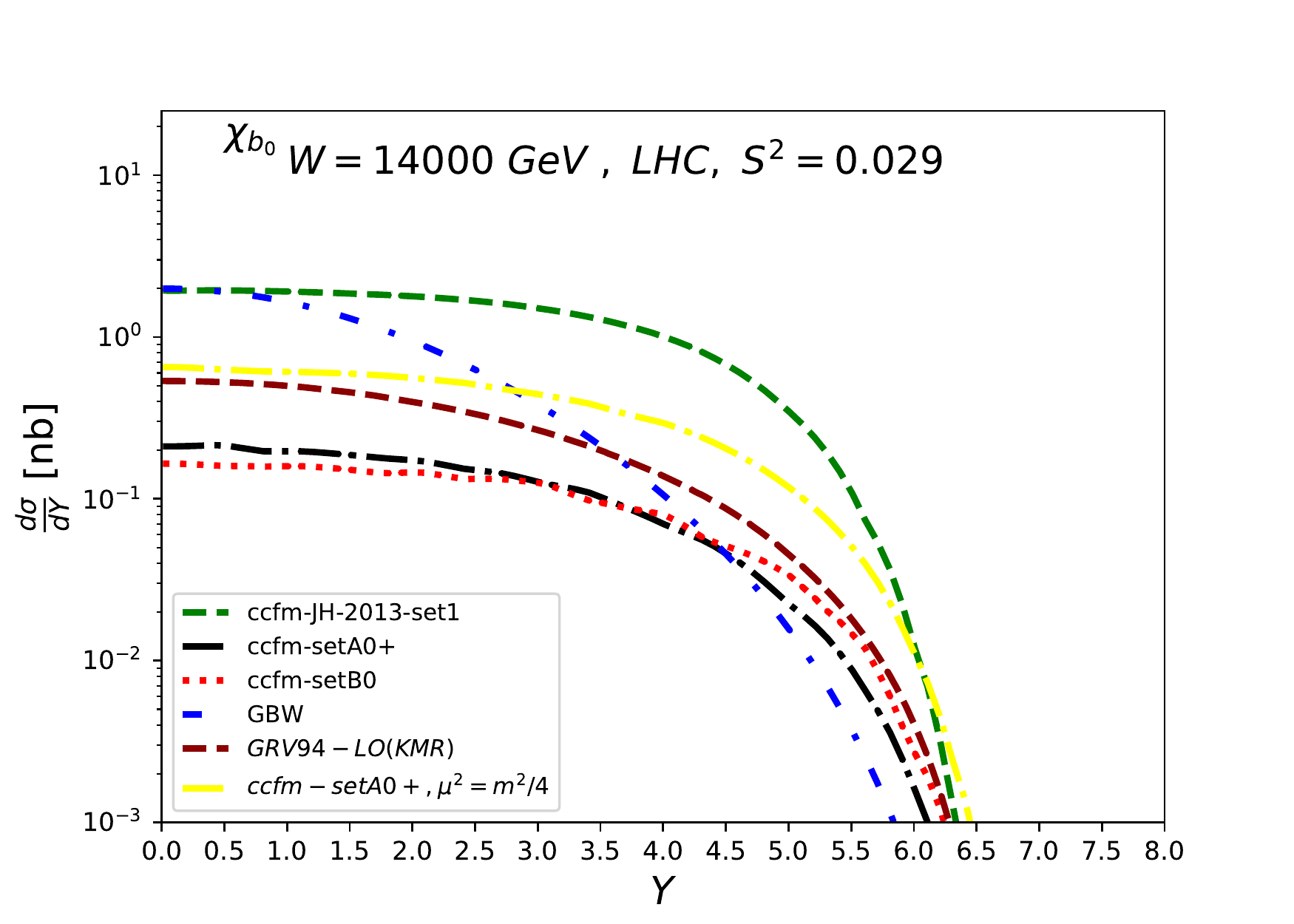} 
	\caption{Rapidity distribution for exclusive $\chi_{b0}$ double diffractive production at LHC using four different sets of UGD's. }
	\label{fig:chiblhc}
\end{figure}

\begin{figure}[t]
		\includegraphics[scale=0.45]{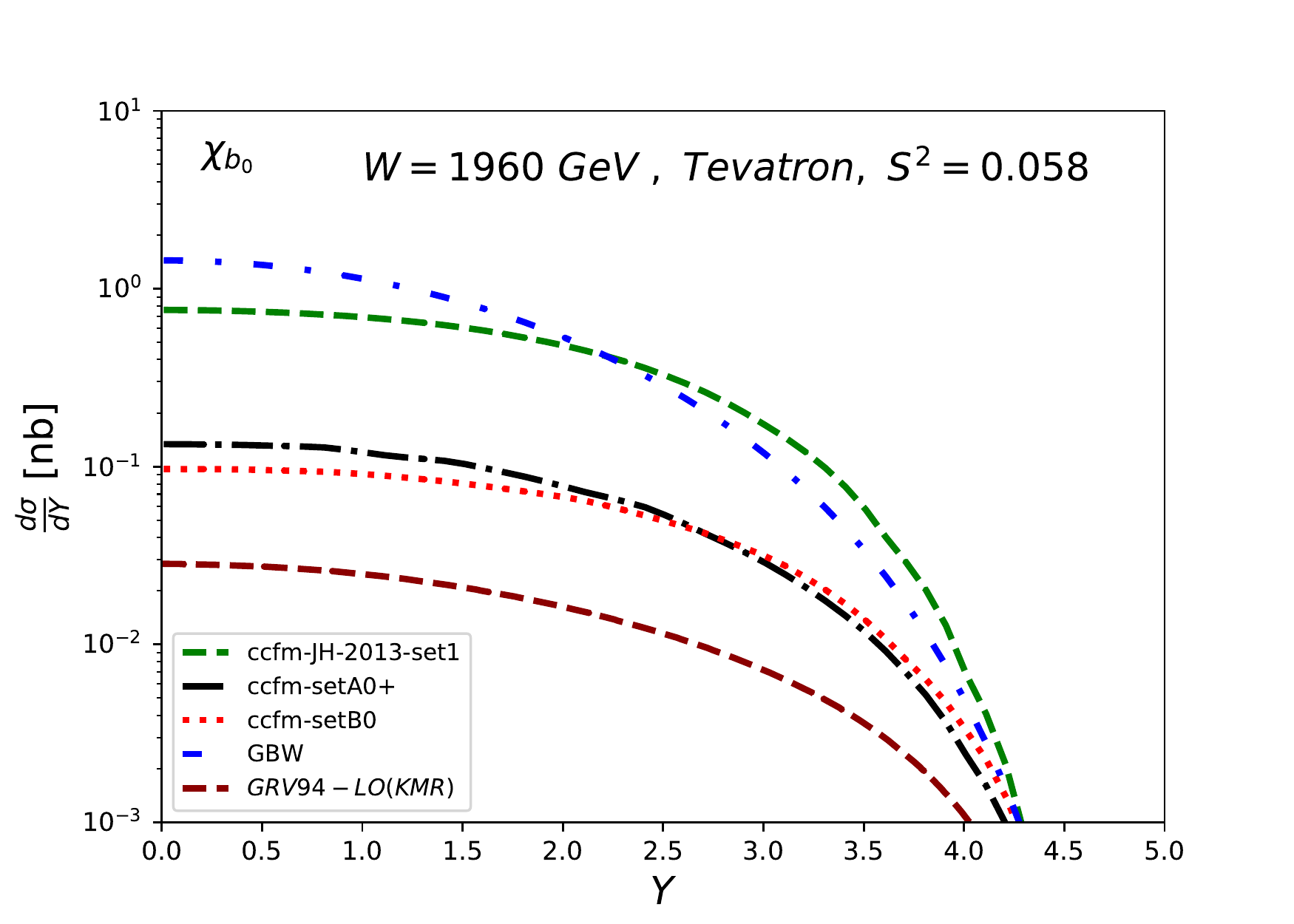} 
	\caption{The exclusive $\chi_{b0}$ double diffractive production using four different sets of UGD's at Tevatron. }
	\label{fig:chibtevatron}
\end{figure}

The evaluations for $\chi_b$ production are presented in Figs. \ref{fig:chiblhc} (LHC) and \ref{fig:chibtevatron} (Tevatron). In both cases the cross section normalization is strongly dependent on the chosen UGD. Moreover, it is verified that the sensitivity to the hard scale $\mu^2$ is not so strong in the rapidity distribution. This is shown in Fig. \ref{fig:chiblhc} for the CCFM set A0+ at LHC energy.  One has $\frac{d\sigma}{dy}(y=0)\sim 100$ pb for Tevatron and $\frac{d\sigma}{dy}(y=0)\sim 500$ pb with $R_g=1$ and a sizable spread for LHC case. It is clear from the present investigation that the main source of uncertainty in the calculations  comes from the model for the UGD.

Interestingly, the GBW UGD allows us to obtain an analytical expression for the rapidity distribution.   Defining an effective saturation scale, $\bar{Q}_s^2 = Q_s^2(x_1)Q_s^2(x_2)/(Q_s^2(x_1)+ Q_s^2(x_2))$, and using the analytical expression in Eq. (\ref{GBWUGD}) one has for their product the following:
\begin{eqnarray}
 {\cal{F}} (x_1,k_{\perp}){\cal{F}} (x_2,k_{\perp}) &=& \left(\frac{3\sigma_0}{4\pi^2\alpha_s}\right)^2\frac{k_{\perp}^6}{[Q_s^2(x_1)+ Q_s^2(x_2)]} \nonumber \\
 &\times& \left(\frac{k_{\perp}^2}{\bar{Q}_s^2}\right)\exp\left(-\frac{k_{\perp}^2}{\bar{Q}_s^2}, \right) 
 \label{prodgbws}
\end{eqnarray}
where the effective saturation scale tends to $\bar{Q}_s^2\approx {Q}_s^2(x_2)$ at large backward rapidities whereas $\bar{Q}_s^2\approx {Q}_s^2(x_1)$ at large forward rapidities. Moreover, at central rapidity one has $\bar{Q}_s^2= {Q}_s^2(x)/2$ where $x=x_1=x_2$. In Eq. (\ref{prodgbws}) one verifies an approximate scaling behavior on the ratio $\tau=k_{\perp}^2/\bar{Q}_s^2$ and then we can rewrite the transverse momentum integration of Eq. (\ref{xs-ch}) in the following form: 
\begin{eqnarray}
 \int_{0}^{\infty}  \frac{dk_{\perp}^2}{k_{\perp}^4} \frac{{\cal{F}} (x_1){\cal{F}} (x_2)}{(m_{\chi}^2+k_{\perp}^2)^2} =  \frac{A\,(\bar{Q}_s^2/m_{\chi}^2)^2\,I_s}{[Q_s^2(x_1)+ Q_s^2(x_2)]}, 
 \end{eqnarray}
 where, the remaining integral is given by,
 \begin{eqnarray}
  I_s(\bar{Q}_s^2,m_{\chi}^2)=\int_{0}^{\infty} \left[\frac{\tau}{1+(\tau/\xi)} \right]^2\frac{d\tau}{e^{\tau}},
\end{eqnarray}
where $A=(3\sigma_0/4\pi^2\alpha_s)^2$ and $\xi = m_{\chi}^2/\bar{Q}_s^2$. The integration over $\tau$ can be explicitly done, which reads as $I_s=e^{\xi}(2\xi^2+\xi^3) Ei_2(\xi)-\xi^2$ with $\xi >> 1$ for the values of $m_{c,b}$. By using the leading terms in the asymptotic expansion of the exponential integral function, $Ei_2(\xi)\approx \frac{e^{-\xi}}{\xi}[1-(2/\xi)+(6/\xi^2)+\ldots]$, an approximate expression for rapidity distribution can be obtained. In the complete case the rapidity distribution is given as,
\begin{eqnarray}
\frac{d\sigma}{dy}\approx 
 \frac{\pi^4\alpha_s^2}{B^2}\frac{A^2m_{\chi}|R'_P(0)|^2}{[Q_s^2(x_1)+ Q_s^2(x_2)]^2}\left(\frac{\bar{Q}_s^2}{m_{\chi}^2} \right)^4I_s^2.
 \label{gbwfull}
\end{eqnarray}

By writing down the expression above in terms of energy and rapidity, one obtains,
\begin{eqnarray}
 \frac{d\sigma}{dy}\approx N_0\left( \frac{\sqrt{s}}{m_{\chi}}\right)^{2\lambda}\mathrm{sech}^6(\lambda y),
\end{eqnarray}
with an overall normalization given by $N_0= S^2R_g^4(\frac{\pi^2\alpha_sAx_0^{\lambda}I_s}{8B m_{\chi}^2})^2 (|R'_P(0)|^2/m_{\chi}^3)$. Here, it is considered $\alpha_s=0.335$ and $\alpha_s=0.25$ for $\chi_c$ and $\chi_b$, respectively.

As a cross check of evaluation of Eq. (\ref{gbwfull}) (with $R_g=1$), one obtains $\frac{d \sigma^{\mathrm{theo}}_{\chi_{c0}}}{dy}(y=0)= 77$ nb for Tevatron, which is consistent with the measured value $(76\pm 14)$ nb \cite{Aaltonen:2009kg}. Also, LHCb have reported preliminary results on exclusive $\chi_{c}$ meson production in the $\chi_c\to J/\psi\,+\,\gamma$ channel~\cite{LHCb:2011dra}, in the rapidity kinematic region $2.0<\eta<4.5$. The cross section times branching ratios (taken from PDG \cite{PDG2019}) for production in the LHCb acceptance ($\varepsilon_s=0.76$) given
by saturation model for $\chi_{c0}$ is 29 pb with large uncertainty compared to the measured value $9.3\pm 4.5$ pb. It can be noticed that the $\chi (J=1,2)$ production amplitudes are identically zero in the perturbative two-gluon exchange model we are using. However, by considering the normalization of $gg\rightarrow \chi_J$ and the mass difference, it is estimated that the cross sections for those states could be a factor $\sim$ 0.7 and 0.06 times the cross section fo $J=0$ state. This gives 20.3 pb and 1.74 pb, compared to experimental values  $16.4 \pm 7.1$ pb and $28\pm 12.3$ pb, respectively.  For comparison, the corresponding prediction from SuperCHIC \cite{Harland-Lang:2015cta} is 14 pb, 9.8 pb and 3.3 pb, respectively.

Finally, the predictions for quarkonium CEP cross sections at different collider energies are considered.  In Table~\ref{tab:1}  the differential cross sections for the central exclusive $\chi_c$ (and $\chi_{b0}$)  production at RHIC, Tevatron and LHC energies are shown. It was verified that the predictions are a factor 2 higher than those from the Durham model for $\chi_{c0}$ \cite{Harland-Lang:2014lxa}. 

\begin{table}[t]
\begin{center}
\caption{Differential cross section (in nb) at rapidity $y=0$ for central exclusive $\chi_{c0}$ and  $\chi_{b0}$  production at RHIC ( at 500 GeV), Tevatron and LHC energies using the saturation model for the UGD. The prompt production $J/\psi\gamma$ and $\Upsilon\gamma$ are also presented.}
\begin{tabular}{|l|c|c|c|c|c|c|}
\hline
$\sqrt{s}$ (TeV)&0.5&1.96&7& 8& 13& 14\\
\hline
$\frac{d\sigma}{dy}(\chi_{c0})$&66 & 77 & 87 & 87.4 & 91.4 & 91 \\
\hline
$\frac{d\sigma}{dy}(\chi_{b0})$&1.27  & 1.6 & 1.9 & 1.94 & 2.08 & 2.1 \\
\hline
$\frac{d\sigma}{dy}(J/\psi\gamma)$&3.65  & 4.53 & 5.44 & 5.50 & 6.00 & 6.01\\
\hline
$\frac{d\sigma}{dy}(\Upsilon\gamma)$&0.113  & 0.14 & 0.16 & 0.17 & 0.18 & 0.19 \\
\hline
\end{tabular}
\label{tab:1}
\end{center}
\end{table}

The perturbative two-gluon exchange model can also be used to compute the prompt production of $V=J/\psi,\Upsilon$ in the process $p+p(\bar p)\rightarrow p+V\gamma+p(\bar p)$. The CEP cross section for this channel is given by  \cite{Yuan:2001nu}, 
\begin{eqnarray}
\label{xs-Vg}
\frac{d\sigma}{dy_{\gamma}dy_{V}d^2p_{\perp}} &= & S^{2}\frac{2\pi^2\alpha_s^2\alpha_{em}e_q^2 m_{V}}{B^2}|R_S(0)|^2\left|\frac{I_V}{m_{\perp}^2x_1x_2s} \right|^2, \nonumber\\
I_V & = & \int \frac{dk_{\perp}^2}{(k_{\perp}^2)^2}{\cal{F}}_g(x_1,x_1',k_{\perp}){\cal{F}}_g(x_2,x_2',k_{\perp}), 
\end{eqnarray}
where $y_{\gamma}$ and $y_{V}$ are the photon and meson  rapidities. The meson transverse momentum is denoted by $\vec{p}_{\perp}$ with a transverse mass $m_{\perp}=\sqrt{m_V^2+p_{\perp}^2}$. Now,  $x_1=\frac{m_{\perp}}{\sqrt{s}}e^{y_{V}} + \frac{p_{\perp}}{\sqrt{s}}e^{y_{\gamma}}$ and $x_2=\frac{m_{\perp}}{\sqrt{s}}e^{-y_{V}} + \frac{p_{\perp}}{\sqrt{s}}e^{-y_{\gamma}}$.  For the masses and radial $S$-wave functions at origin \cite{Eichten:2019hbb}, we use $m(\psi)=3.096$  GeV with $|R_S(0)|^2_c=0.81$ GeV$^5$ and $m(\Upsilon)=9.46$ GeV with $|R_S(0)|^2_b=6.48$ GeV$^5$. Once again, the saturation model gives an analytical solution for the integral $I_V$. Therefore, the differential cross section  is written as,
\begin{eqnarray}
\label{xs-VgGBW}
\frac{d\sigma}{dy_{\gamma}dy_{V}d^2p_{\perp}} &= & S^{2}\frac{8\pi^2\alpha_s^2\alpha_{em}e_q^2 m_{V}A^2}{B^2(sx_1x_2m_{\perp}^2)^2}|R_S(0)|^2 \nonumber \\ 
&\times& \frac{(\bar{Q}_s^2)^4}{[Q_s^2(x_1)+ Q_s^2(x_2)]^2}. \nonumber 
\end{eqnarray}

\begin{figure}[t]
		\includegraphics[scale=0.45]{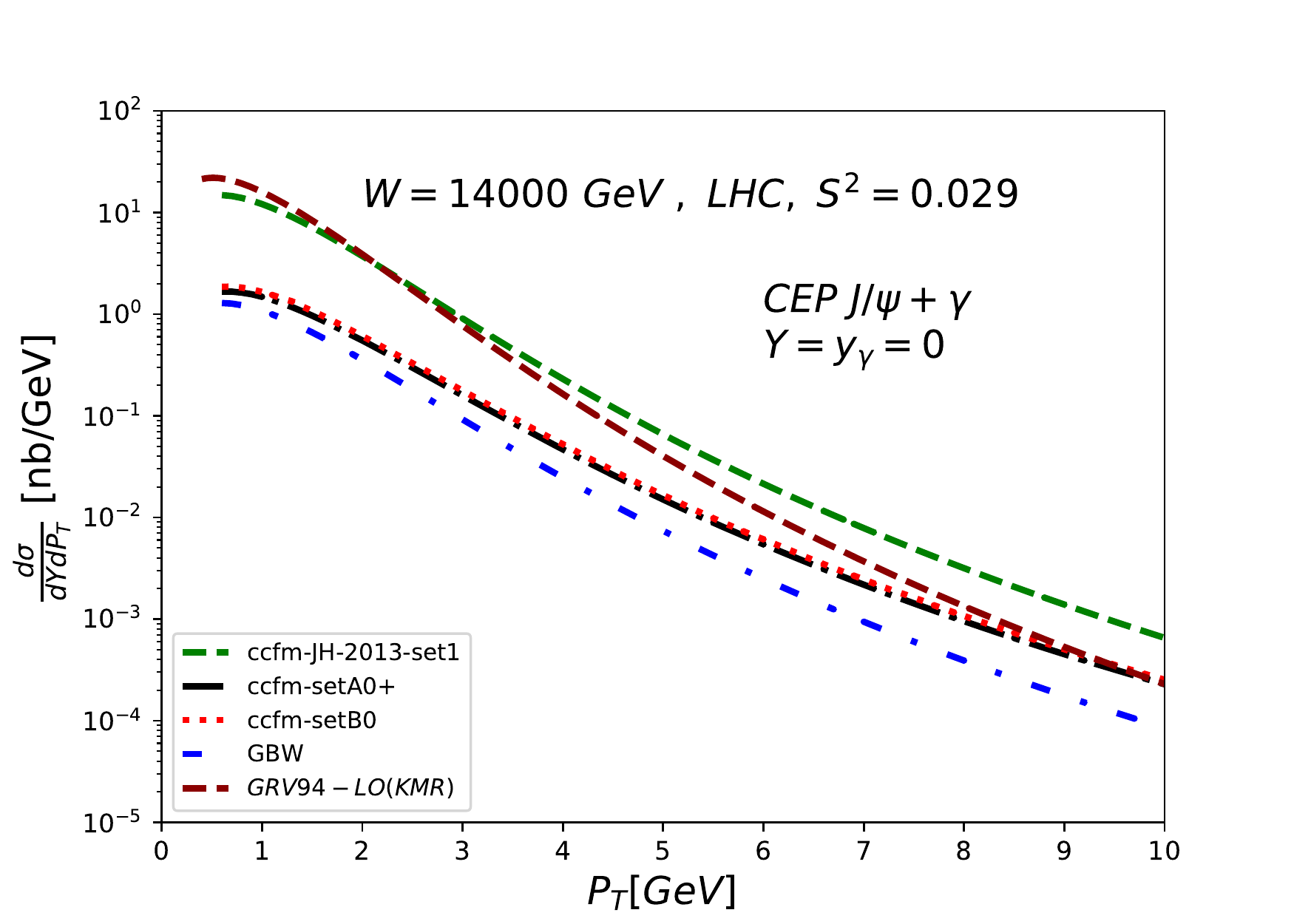} 
	\caption{Differential cross section for process $p+p\rightarrow p+J/\psi \gamma +p$ at the LHC as a function of mesons transverse momentum at $y_{\psi}=0$. }
	\label{fig:psigammalhc}
\end{figure}

\begin{figure}[t]
		\includegraphics[scale=0.45]{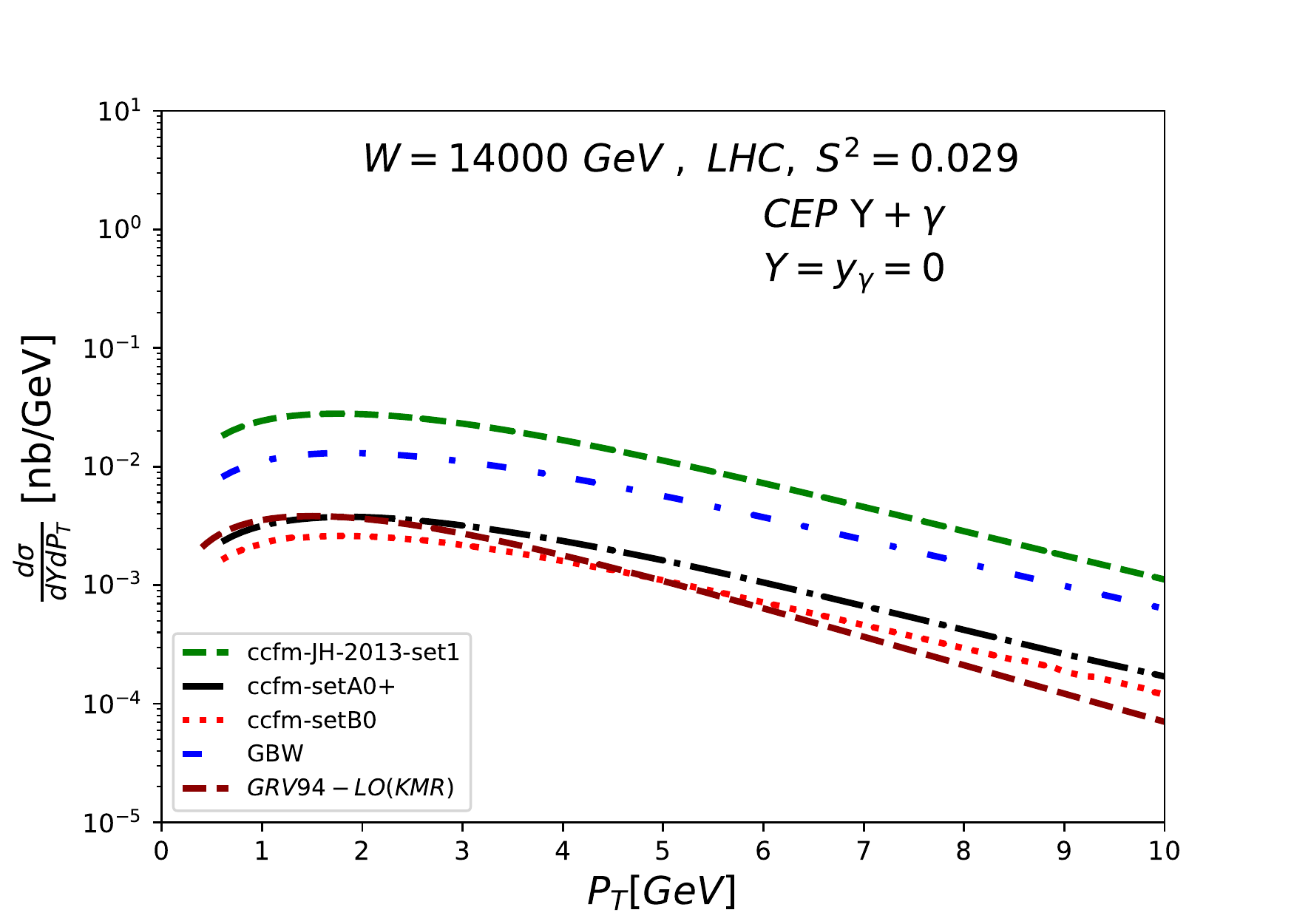} 
	\caption{Differential cross section for process $p+p\rightarrow p+\Upsilon \gamma +p$ at the LHC as a function of mesons transverse momentum at $y_{\Upsilon}=0$. }
	\label{fig:upsilongammalhc}
\end{figure}

The numerical calculation for the differential cross sections   for production of $J/\psi +\gamma$ and $\Upsilon+\gamma$ at central rapidity are presented in Table~\ref{tab:1} (integrated over photon rapidity and meson transverse momentum).  The cross sections for S-wave
quarkonia are comparable or larger than those for P-wave states times $\mathrm{Br}(\chi \rightarrow V\gamma)\sim 10^{-2}$ at least at $y=0$. This is  in disagreement with the conclusions presented in Ref. \cite{Yuan:2001nu}, which predicts that the leading contributions to CEP of S-wave quarkonium are the feeddown contributions from P-wave decays. In Figs. \ref{fig:psigammalhc} and \ref{fig:upsilongammalhc} we present the differential cross sections, Eq. (\ref{xs-VgGBW}) integrated over photon rapidities, in terms of meson transverse momentum at $y_V=0$ for the differents UGDs discussed before. We found that the main contribution for the meson $p_{\perp}$-spectra comes from the region $p_{\perp}\lesssim m_V$. 

Before discussing the integrated cross sections for different models of UGDs, one needs to estimate the extrapolation of the saturation model to nuclear collisions. It is found in Eqs. (\ref{gbwfull}) and (\ref{xs-VgGBW}) that the rapidity distributions take the form $d\sigma/dy \propto (\sigma_0)_1^2(\sigma_0)_2^2[\bar{Q}_s^2(x_1,x_2)]^4/[Q_s^2(x_1)+Q_s^2(x_2)]^2$. Let us consider the label 1 for projectile and 2 for the target and take into account the geometric scaling property in nuclear reactions demonstrated in Ref. \cite{Armesto:2004ud}. Namely, for the unintegrated gluon distribution in a nucleus we could replace in Eq. (\ref{GBWUGD}) $\sigma_0^A=(\pi R_A^2/\pi R_p^2)\sigma_0\sim A^{2/3}\sigma_0$ and $Q_{s,A}^2(x) =[A\pi R_p^2/\pi R_A^2]^{\frac{1}{\delta}}Q{s}^2(x)\sim A^{1/3\delta}Q_{s}^2(x)$. Here, $\delta=0.79$ and $Q_s(x)$ is the saturation scale for the proton case. Explicitly for $pA$ collisions it gives:
\begin{eqnarray}
\frac{d\sigma_{pA}}{dy} &\propto& (\sigma_0)_p^2(\sigma_0)_A^2\left\{ \frac{(Q_s^2(x_1))^4(Q_{s,A}^2(x_2))^4}{[Q_s^2(x_1)+Q_{s,A}^2(x_1)]^6} \right\} \nonumber \\
&\approx & \left(\frac{\pi R_A^2}{\pi R_p^2}\right)^2\left(\frac{A\pi R_p^2}{\pi R_A^2} \right)^{-2/\delta} \frac{d\sigma_{pp}}{dy} ,
\end{eqnarray}
and similarly for $AA$ collisions,
\begin{eqnarray}
\label{Adep}
\frac{d\sigma_{AA}}{dy} &\propto& (\sigma_0)_A^2(\sigma_0)_A^2\left\{ \frac{(Q_{s,A}^2(x_1))^4(Q_{s,A}^2(x_2))^4}{[Q_{s,A}^2(x_1)+Q_{s,A}^2(x_1)]^6} \right\} \nonumber \\
&\approx & \left(\frac{\pi R_A^2}{\pi R_p^2}\right)^4\left(\frac{A\pi R_p^2}{\pi R_A^2} \right)^{2/\delta} \frac{d\sigma_{pp}}{dy}.
\end{eqnarray}

The crude approximation above based on the geometric scaling property can be compared to the sophisticated calculations using SuperCHIC 3 Monte Carlo \cite{Harland-Lang:2018iur}, which implements CEP in nuclear collisions.

\subsection{Integrated cross section}
Based on the rapidity distribution obtained above, the integrated cross section can be computed. Results for Tevatron and LHC energies are shown in Tab. \ref{tab:2}. The output for the different UGS's sets are presented for $\chi_{c0}$ ($\chi_{b0}$) in units of nanobarns and disregarding the skewedness effects. At the LHC, the larger cross section corresponds to the GRV94-LO UGD, whereas GBW UGD gives lowest values. Based on the theoretical uncertainty from UDG's  one obtains $\sigma(\chi_{c0})=3619 \pm 3241$ nb and  $\sigma(\chi_{b0})=4.7\pm 3.5$ nb at LHC with $R_g=1$. 

In case of $\chi_{c,b}$ to be measured by detecting their radioactive decays to quarkonium plus photon the final cross section for quarkonium production from $\chi_0$ feeddown decays would be:
\begin{eqnarray}
\frac{\sigma (\chi_{c0})}{R_g^4}\mathrm{Br}(\chi_{c0}\rightarrow J/\psi\gamma)\mathrm{Br}(J/\psi\rightarrow \mu^+\mu^-)\approx   1.41\pm 1.11\, \mathrm{nb}, \nonumber
\end{eqnarray}
for $\chi_{c0}$ exclusive production. On the other hand, for the $\chi_{b0} $ production one obtains,
\begin{eqnarray}
\label{brachchib}
& &\frac{\sigma (\chi_{b0})}{R_g^4}\left[\mathrm{Br}(\chi_{b0}(2P)\rightarrow \Upsilon(2S)\gamma)\mathrm{Br}(\Upsilon(2S)\rightarrow \mu^+\mu^-)  \right. \nonumber \\
& & + \left. \mathrm{Br}(\chi_{b0}(2P)\rightarrow \Upsilon(1S)\gamma)\mathrm{Br}(\Upsilon(1S)\rightarrow \mu^+\mu^-) \right] \nonumber \\   & &\approx  1.9\pm 1.45\, \mathrm{pb},\nonumber
\end{eqnarray}
where it has been assumed for simplicity that the production cross section for $\chi_b$ in the $2P$ and $1P$ states are of same order of magnitude.  

\begin{table}[t]
\begin{center}
\caption{Integrated cross sections for Tevatron ($\sqrt{s}=1960$  TeV) and LHC ($\sqrt{s}=14$ TeV). Results for $\chi_{c_{0}}$ ($\chi_{b_{0}}$) are in units of nanobarns (nb) and with R$_{g}=1$.}
\begin{tabular}{|l|c|c|}
\hline
UGD & Tevatron&LHC\\
\hline 
GBW & 294 (5.0) & 378 (8.2)\\
\hline
CCFM-JH2013 & 1452 (3.5) & 4973 (15)   \\ 
\hline
CCFM-setB0 & 840 (0.5) &  795 (1.23)\\ 
\hline
CCFM-setA0+ & 620 (0.6) & 560 (1.4)\\ 
\hline
GRV94-L0  & 551 (0.13)  & 6860 (3.2) \\ 
\hline
\end{tabular}
\label{tab:2}
\end{center}
\end{table}

We now compare our results to other models available in the literature. The SuperCHIC MC generator implements the Durham model and the $\chi_c$ cross sections according to it at $\sqrt{s} = 7$ TeV, over the full kinematic
range and including the branching ratios of $\chi_c \rightarrow J/\psi\gamma \rightarrow \mu^+\mu^-$ are 194 pb,
133 pb and 44 pb for $\chi_{c0}$, $\chi_{c1}$ and $\chi_{c2}$ respectively. The predictions in this work are considerably larger than SuperCHIC \cite{Harland-Lang:2015cta}, with the saturation model being the closest one ($\approx 300$ pb). The measured value by LHCb is $\simeq 134$ pb. Interestingly, high cross sections were also obtained in Ref. \cite{Rangel:2006mm}, using Bialas-Landshoff (BL) formalism implemented in DPEMC Monte carlo. The BL model was also applied to $\chi$ production in Ref. \cite{Bzdak:2005rp}, with a cross section of 350 nb for
$\chi_{c0}$ production at the LHC.  Predictions are also consistent in order of magnitude with results presented by Cracow/Lund group in Ref. \cite{Pasechnik:2007hm}, including the large uncertainty from the choice for the UGDs. The same occurs for results from 3-Pomeron model \cite{Ryutin:2012np}, which predicts $\sigma(\chi_{c0})= 212\pm 53$ nb at 7 TeV (future version of ExDiff Monte Carlo \cite{Ryutin:2018per}, based on theoretical framework of Ref. \cite{Ryutin:2012np} will include quarkonium production).

The predicted $\chi_{b0}$ cross section is much higher than Durham group, probably due to a different coupling of two gluons to the $\chi_b$. The non-perturbative two-gluon model   (BL) from Ref. \cite{Bzdak:2005rp} predicted a total $\chi_{b0}$ cross section of 0.3 nb  at $\sqrt{s} = 14$ TeV, which it is consistent with present calculations using CCFM-setB0 and CCFM-setB0+ or  CCFM-setA0 . Moreover, in Ref. \cite{Yuan:2001nu} was predicted a total cross section of 0.88 nb at the Tevatron, in agreement in order of magnitude with present work. A Regge-eikonal approach for CEP is investigated in Ref. \cite{Petrov:2004nx}, which predicts $\sigma(\chi_{b0}) \simeq 0.16$ nb and 1.3 nb at Tevatron and LHC, respectively. Once again, results presented in Table \ref{tab:2} are consistent with those calculation. 

As final comment, besides being considered the theoretical uncertainties on UGDs and hard scales, other quantities are source of additional uncertainty as the slope of the proton form factor, $B$, the gap survival factor and value of the wave-function at the origin.
The main shortcoming of the present approach is that the higher spins $J=1$ and $2$ contributions are vanishing. This is traced back to the scattering amplitudes for those processes. Namely, writing down the amplitude ${\cal{M}}$ in terms of the $g^*g^*\rightarrow \chi_J$ coupling, $V_J$ \cite{Yuan:2001nu},
\begin{eqnarray}
\label{Mampl}
{\cal{M}}(\chi_J) = \frac{9\pi^2}{4}\int  \frac{dk_{\perp}^2}{k_{\perp}^4} {\cal{F}}_g(x_1,k_{\perp}){\cal{F}}_g(x_2,k_{\perp}) V_J, 
\end{eqnarray}
it can be demonstrated that the $\vec{k}_{\perp}$ integration above gives values equal zero. For the $J=1$ state, which has a polarization vector $\epsilon^{\mu}_{(J=1)}$ one has 
\begin{eqnarray}
V_1\propto \frac{|R'_P(0)|^2}{(m_{\chi}^2+k_{\perp}^2)^2}\varepsilon_{\mu \nu \rho \sigma} \epsilon^{\mu}_{(J=1)}k_{\perp}^{\nu}p_1^{\rho}p_2^{\sigma},
\end{eqnarray}
with $V_1\propto k_{\perp}^{\nu}$ and the corresponding amplitude will be zero after angular integration of Eq. (\ref{Mampl}). The situation is more involved for the $J=2$ state, which has a polarization tensor denoted by $\epsilon^{\mu \nu}_{(J=2)}$,  obeying both properties $\epsilon_{\mu\nu}P^{\mu}=0$ and $\epsilon_{\mu \nu}g^{\mu \nu} =0$. From direct inspection of the coupling for this state,
\begin{eqnarray}
\label{V2}
V_2 &\propto &  \frac{m_{\chi}^2|R'_P(0)|^2}{k_{\perp}^2(m_{\chi}^2+k_{\perp}^2)^2}\epsilon^{\mu \nu}_{(J=2)} \\
&\times & \left[4k_{\perp}^2\left(p_{1\mu}p_{2\nu}+ p_{1\nu}p_{2\mu} \right) +s\left(P_{\mu}P_{\nu} - 4[k_{\perp}]_{\mu}[k_{\perp}]_{\nu} \right)\right], \nonumber
\end{eqnarray}
one concludes that after integrating the azimuthal angle of $\vec{k}_{\perp}$, where $\int d^2kk^{\mu}k^{\nu}=(\pi/2)\int dk^2 k^2 g_{\mu\nu}^{(T)}$, the expression in the brackets in the  second line of Eq. (\ref{V2}) becomes $s(P_{\mu}P_{\nu} - 2g_{\mu\nu}k_{\perp}^2)$. 
This implies that the amplitude for $J=2$  will be equal to zero due to the properties of the polarization tensor. Here, $ g_{\mu\nu}^{(T)}$ is the transverse component of the tensor  $ g_{\mu\nu}$. 

A vanishing contribution to $J=1,2$ states is also shared by models of Refs.  \cite{Bzdak:2005rp,Rangel:2006mm,Ryutin:2012np} in the very forward limit. Non vanishing contributions are obtained for different coupling prescriptions. For soft Pomeron models, where the Pomeron couples like an even charge conjugation object (similar to photon) the $\chi_J$ production amplitude has a coupling analogous to the process of $\gamma^*\gamma^*\rightarrow \chi_J$ \cite{Kuhn:1979bb}. In general, this prescription leads to similar magnitude production rates for states $J=1,2$ compared to $J=0$ state (see, e.g. Ref. \cite{Machado:2011vh}). This procedure is behind the recent calculations of the Durham Group for exclusive $\chi_J$ production and they used the formalism of Kuhn et al. \cite{Kuhn:1979bb} for the first time in Ref. \cite{HarlandLang:2009qe}. The Cracow/Lund Group has proposed a general expression for the coupling of the two virtual gluons to the $\chi_J$-meson based on the quasi-multi-Regge-kinematics (QMRK) approach. For the axial-vector ($J=1$) quarkonia, it was shown in \cite{Pasechnik:2009bq} that a non-vanishing amplitude is obtained for off-shell gluons and the interplay between the off-shell matrix element and off-diagonal UGDFs has been discussed. Afterwards, the analysis for the tensor $\chi (J=2)$ meson was done in Ref. \cite{Pasechnik:2009qc}, showing that a relative suppression on axial-vetor meson production with respect to scalar and tensor ones implies to the dominance of the $\chi(J=2)$ contribution over the $\chi (J=1)$ one in the radiative decay channel. In that same work, authors demonstrated that their results for the hard subprocess amplitudes are in full agreement with the corresponding results from the Durham Group \cite{HarlandLang:2009qe}. Therefore, both groups predict roughly a smaller rate from the axial-vector meson compared to the tensor one. Of course, the number of uncertainties coming from distinct kinematical cuts and various models for the UGDs makes a direct comparison a complex task. In any case, the rates for $J=1,2$ mesons are somewhat model dependent since they are based on the analogy with the process $\chi_J\rightarrow \gamma^* \gamma^*$ (an analysis along these lines for inclusive $\chi_{c,b}(0^+)$ production was done recently in Ref. \cite{Babiarz:2020jkh}). For example, for  Tevatron energy the Cracow/Lund group predicts the ratios $\chi_c(1^+)/\chi_c(0^+)=0.1 (0.1)$ and  $\chi_c(2^+)/\chi_c(0^+)=0.4\,(0.3)$ [for KS UDG (KMR UGD )] whereas the Durham Group preditcs $\chi_c(1^+)/\chi_c(0^+)=0.8\,(0.6)$ and  $\chi_c(2^+)/\chi_c(0^+)=0.6\,(0.2)$ based on Ref. \cite{HarlandLang:2009qe} (Ref. \cite{Harland-Lang:2014lxa}). Both calculations contain very large theoretical uncertainties and the amount of
 $\chi (2^+)$ experimentally observed \cite{LHCb:2011dra} seems to be larger than predicted. The results from Refs. \cite{Pasechnik:2009qc,Harland-Lang:2014lxa,HarlandLang:2009qe} can be directly compared to our results in Table \ref{tab:1} for $\chi (0^+)$ (see, e.g. Table I of Ref.  \cite{Pasechnik:2009qc} and Tables 2-3 of Ref. \cite{Harland-Lang:2014lxa}) , whereas our predictions for $\chi_b(0^+)$ have been discussed in this subsection (Durham Group provides $\chi_b(1^+)/\chi_b(0^+)\approx 0.055$ and  $\chi_b(2^+)/\chi_b(0^+)\approx 0.14$). It is worth mentioning that in the present work the prompt $J/\psi+\gamma$ and $\Upsilon+\gamma$ production is predicted for the first time for LHC energies using the very same formalism as for the $\chi$ states.

\section{Summary}

We have investigated the central exclusive production of $\chi_{c,b0}$ in hadron-hadron collisions. In the theoretical calculations, it was taken into account the perturbative two-gluon model and non-relativistic approximation for meson wave functions. The numerical results are obtained for different models for the unintegrated gluon distribution, including an analytical parametrization from parton saturation approach. It was found that the main uncertainty in the prediction comes from the choice for the UGD, and verified that the different prescriptions for the hard scale $\mu^2$ have a small effect for $\chi_b$ production. It was also shown that the saturation model for the UGD allow us to obtain an analytical expression for the rapidity distribution both for $\chi_J$ and prompt production $V\gamma$. It depends explicitly on the effective saturation scale, $\bar{Q}_s(x_1,x_2)$, and can be easily extended to $pA$   or $AA$ collisions using arguments of geometric scaling. That is, the nuclear saturation scale is rescaled compared to the nucleon one, $Q_{s,A}^2\propto A^{1/3}Q_{s,p}^2$. We found that the corresponding scaling is $\sigma_{coh}^{pA}\sim A^{\alpha_{pA}}\sigma_{pp}$ for proton-nucleus (with $\alpha_{pA}=(4\delta-2)/3\delta\approx 4/9$)   and  $\sigma_{coh}^{AA}\sim A^{\alpha_{AA}}\sigma_{pp}$ (with $\alpha_{AA}=(8\delta+2)/3\delta)\approx 10/3$) for nucleus-nucleus collisions, respectively.

Summarizing the results from the analytical expressions based on the saturation model at 7 TeV, one has $\mathrm{Br}\times\sigma_{\chi_{c0}}(2.0<y<4.5) = 29$ pb, and $\mathrm{Br}\times\sigma_{tot}(\chi_{c0})= 300$ pb. Moreover, one obtains  $\mathrm{Br}\times\sigma_{tot}(\chi_{b0})\simeq 2.2$ pb using the approximations discussed,  Eq. (\ref{brachchib}). Considering the decay channels $\chi_{c0}(1P)\rightarrow K^+K^-$ and $\chi_{c0}(1P)\rightarrow \pi^+\pi^-$ at 8 TeV, we estimate $\mathrm{Br}(K^+K^-)\sigma_{tot}(\chi_{c0})\simeq 123$ pb and  $\mathrm{Br}(\pi^+\pi^-)\sigma_{tot}(\chi_{c0})\simeq 113$ pb (with cut $2.5<y_{\chi}<4.5$). 

Our study demonstrated that the CEP of mesons is a powerful tool to investigate the perturbative QCD dynamics and  in proton-proton collisions at the LHC. This shall stimulate further experimental and theoretical studies.

\begin{acknowledgments}
	We are grateful to Hannes Jung for interesting discussions and comments. This work was  partially financed by the Brazilian funding
	agencies CNPq, CAPES and FAPERGS.
\end{acknowledgments}

\end{document}